# Query-limited Black-box Attacks to Classifiers

**Fnu Suya**
University of Virginia

**Yuan Tian**
University of Virginia

**David Evans**
University of Virginia

**Paolo Papotti**
EURECOM

## Abstract

We study black-box attacks on machine learning classifiers where each query to the model incurs some cost or risk of detection to the adversary. We focus explicitly on minimizing the number of queries as a major objective. Specifically, we consider the problem of attacking machine learning classifiers subject to a budget of feature modification cost while minimizing the number of queries, where each query returns only a class and confidence score. We describe an approach that uses Bayesian optimization to minimize the number of queries, and find that the number of queries can be reduced to approximately one tenth of the number needed through a random strategy for scenarios where the feature modification cost budget is low.

## 1 Introduction

Recent works reveal the vulnerabilities of current machine learning models to carefully-crafted adversarial examples [1, 2, 3, 4]. In many scenarios, complete model information is not available to the attacker and hence it is important to study black-box attacks, where the attackers do not have full knowledge of the model but only some way of interacting with it as a black box. In this work, we focus on attacks where only query access to the model is available and each query result consists of a predicted class and confidence score.

Since queries to the model are costly, attackers are motivated to minimize the number of queries needed when interacting with the model. For example, if the adversary is attempting to craft messages that evade a spam email detection systems, each query to the underlying classification model involves sending an email and possibly poisoning an account, so the adversary is not be able to afford large number of email queries [5]. Hence, our problem is: find an adversarial example that satisfies the constraints on feature modification with the minimal number of queries. This problem can be cast as a constrained optimization problem:

$$\min Q(\mathbf{x}) \quad \text{s.t.} \ f(\mathbf{x}) \neq f(\mathbf{x}^A) \wedge c(\mathbf{x}, \mathbf{x}^A) \leq C \quad (1)$$

where $Q(\mathbf{x})$ denotes the total number of queries consumed in searching for an adversarial example for seed sample $\mathbf{x}$, $f(\mathbf{x})$ denotes the prediction label of instance $\mathbf{x}$, $c(\mathbf{x}, \mathbf{x}^A)$ denotes feature modification cost, where $\mathbf{x}^A$ is the original instance, and $C$ is the budget for feature modification. In this paper, we use the $L_1$-norm to measure cost, $c(\mathbf{x}, \mathbf{x}^A) = ||\mathbf{x} - \mathbf{x}^A||_1$, as is commonly used in the text domain.

This optimization problem is highly intractable as we do not have a closed form expression for function $Q(\mathbf{x})$. Further, $f(\mathbf{x})$ is unknown since we only have black-box access to the machine learning model. Because of high intractability of the problem in Eq. (1), it is important for us to apply some transformations to make the whole problem tractable. In particular, we convert the constrained optimization form into one amenable to Bayesian optimization (Section 3).

The main contributions of our work are introducing a new formulation for black-box adversarial machine learning where minimizing query numbers and proposing a query-minimizing black-box attack strategy based on Bayesian optimization. Section 4 reports on preliminary experiments using these techniques to generate spam messages that evade a black-box detector.

**Related Work** Prior works have studied black-box attacks on machine learning classifiers in two categories: *substitute model attacks* and numerical *approximation method-based attacks*. The first type of attack uses query responses obtained from the target model to train a substitute model, and then generates adversarial examples for that substitute model. Adversarial examples produced this way are transferable and often effective against the original model [6, 7, 8]. The drawback of the substitute model is it will suffer form the transfer loss as not all adversarial examples can transfer from one model to another model [9]. Also, the number of training instances needed to produce an effective substitute model may be very large. One recent work in [10] adopts slightly different strategy by learning a separate attacker model, which is trained to produce adversarial samples to the target model. However, the query number minimization is still not explicitly considered in [10].

Another line of work, introduced by Chen et al. [9], is to apply some numerical approximation to model gradient calculation to support known white-box attack strategies. This approximates gradient information by the symmetric difference quotient and further utilizes the Carlini and Wagner attack [11] to generate adversarial examples. This approach requires a large number of queries since the gradient needs to be calculated in each step and each gradient estimation requires many model evaluations because of the high-dimensional feature space.

Previous papers on black-box attacks do not explicitly consider minimizing the total number of model interactions. A closely related work by Li and Vorobeychik [5] considers the spam email setting and sets a bound on the total number of queries and feature modification cost. The attacker then applies a query strategy to find adversarial examples. However, this work only considers linear classifiers. In contrast, our work applies to arbitrary (continuous) classifiers, including neural networks and other linear models.

## 2 Background on Bayesian Optimization

Bayesian optimization (BO) is a derivative free strategy for global optimization of black-box functions [12, 13, 14]. The Bayesian optimization problem can be formulated as:

$$\min g(\mathbf{x}) \quad \text{s.t. } h(\mathbf{x}) \leq 0.$$

where $g(\mathbf{x})$ is an unknown function and $h(\mathbf{x})$ can either be known or unknown. Unlike traditional optimization algorithms, BO does not depend on gradient or Hessian information; instead, it works by querying the function value of a point in each step of the interactive optimization process [12]. As queries to $g(\mathbf{x})$ are assumed to be costly, the algorithm minimizes the total number of queries used in the whole search process.

Since the objective function is unknown, a *prior* over the function is assumed to be known, e.g., Gaussian prior [15] is commonly used [12, 14, 13]. With the defined priors and current observations, the *posterior probability* of next function value can be defined. With the posterior probability distribution, an *acquisition function*, Acq($\mathbf{x}$) is then defined to capture an *exploration-exploitation* trade-off in determining the next query point [16, 17]. Points with larger Acq($\mathbf{x}$) values are likely to have smaller $g(\mathbf{x})$ values. Thus, we prefer to query points where Acq($\mathbf{x}$) is large. Since the goal for each step is to select a point that maximizes the current acquisition function, the whole optimization process heuristically minimizes number of interactions needed to find a solution. Convergence rate of Bayesian optimization can be referred to [16, 17].

Exploration prefers points where the uncertainty is high, while exploitation prefers points where the objective function value is low (for minimization problems). After each function evaluation, the acquisition function is updated along with the posterior probabilities. For this work, we use the upper confidence bound (UCB) selection criterion in selecting the specific acquisition function type. As we assume the unknown function value $g(\mathbf{x})$ at point $\mathbf{x}$ follows Gaussian distribution, we obtain the closed form expression of the acquisition function (UCB) for point $\mathbf{x}$ as Acq($\mathbf{x}$) = $\mu(\mathbf{x}) + \kappa\sigma(\mathbf{x})$, where $\sigma(\mathbf{x})$ and $\mu(\mathbf{x})$ are variance and mean functions, respectively, at point $\mathbf{x}$ and $\kappa$ is a constant. Brouchu et al.'s tutorial [12] provides more details regarding different types of acquisition functions and closed form expressions for $\mu(\mathbf{x}), \sigma(\mathbf{x})$. Once the query result $g(\mathbf{x}_t)$ of the point $\mathbf{x}_t$ is returned, the BO framework updates its belief about the unknown function distribution and the whole procedure iterates until termination condition is satisfied.

## 3 Minimizing Queries with Bayesian Optimization

As discussed in Section 1, we face two major challenges: no closed form expression for function $Q(\mathbf{x})$ and an unknown constraint in $f(\mathbf{x})$, where only queries to $f(\mathbf{x})$ are allowed. We handle the unknown constraint by following the approach used by Carlini and Wagner [11, 1] by moving the intractable classification label constraint into the objective function. Since we do not know $f(\mathbf{x})$, we transform the constraint of $f(\mathbf{x}) \neq f(\mathbf{x}^A)$ as minimizing the probability of $\mathbf{x}$ having same label with $\mathbf{x}^A$. In order to minimize the total number of queries, as outlined in the objective function of Eq. (1), we adopt a heuristic strategy for minimization. Namely, in each query step, we use our query history to select the apparently best point for solving the optimization problem. Hence, specific to our problem, in each query step, we find the best point for minimizing $\Pr[f(\mathbf{x}) == f(\mathbf{x}^A)]$ and consequently, the whole optimization process eventually minimizes function $Q(\mathbf{x})$ (i.e. the total number of queries). Our query step will terminate once we have found a valid instance whose label is different from $\mathbf{x}^A$. The problem can be mathematically formulated as:

$$\min \Pr[f(\mathbf{x}) == f(\mathbf{x}^A)] \text{ s.t. } c(\mathbf{x}, \mathbf{x}^A) \leq C \qquad (2)$$

To solve the problem in Eq. (2), we adopt the Bayesian optimization framework which is well-suited for solving an unknown function (in our case, $\Pr[f(\mathbf{x}) == f(\mathbf{x}^A)]$) minimization with a minimal number of queries (i.e., minimizing $Q(\mathbf{x})$). Note that $c(\mathbf{x}, \mathbf{x}^A)$ is a function known to the adversary (i.e., $L_1$-norm constraint). We now have a Bayesian optimization problem with an unknown objective and known constraint. We use Upper Confidence Bound (UCB) as the acquisition function (Acq($\mathbf{x}$)) and in each step we select the point that maximizes Acq($\mathbf{x}$) with respect to the constraint $c(\mathbf{x}, \mathbf{x}^A) \leq C$.

We apply the DIRECT algorithm [18] to solve acquisition function maximization problem in Eq. (2). DIRECT is a well-known algorithm for solving global optimization problems. To increase the robustness of the code when facing a small cost budget $C$, we applied DIRECT with minor modifications. DIRECT works by dividing a unit hypercube sequentially and evaluating function values in each of the sub-hyper-rectangles [18] and the initial point is center of the unit hypercube. Originally, each dimension value of this point was determined by the lower and upper bounds in that dimension. When $C$ is very small and the initial center is too far away from initial point $\mathbf{x}^A$, it is very hard to find an instance within the feature cost budget (which will result in very long search time). Instead, we now take the initial point $\mathbf{x}^A$ as the center of the unit hypercube such that we can always find instances that satisfy the feature modification cost constraint. Algorithm 1 summaries our Bayesian optimization algorithm; details regarding Guassian process update can be found in Rasmussen [15].

---
**Algorithm 1** Bayesian Optimization Based Black-box Attack
---
**Input:** $\mathbf{x}^A, C, f(\mathbf{x}^A), N$ (maximum number of iterations)
**Output:** $\mathbf{x}^*$
 1: $\mathbf{x}^* = \mathbf{x}^A$
 2: **for** $t = 1, 2, ..., N$ **do**
 3:     Find $\mathbf{x}_t$ by solving problem $\mathbf{x}_t = \text{argmax Acq}(\mathbf{x}|D_{1:t-1})$, s.t. $c(\mathbf{x}, \mathbf{x}^A) \leq C$
 4:     Sample the objective function value: $y_t = \Pr(f(\mathbf{x}_t) == f(\mathbf{x}^A))$
 5:     **if** $f(\mathbf{x}_t) \neq f(\mathbf{x}^A)$ **then**
 6:         **return** $\mathbf{x}^* = \mathbf{x}_t$;
 7:     **end if**
 8:     Augment the data $D_{1:t} = \{D_{1:t-1}, (\mathbf{x}_t, y_t)\}$ and update the Gaussian Process and Acq($\mathbf{x}$).
 9: **end for**
---

## 4 Evaluation

We have conducted preliminary experiments to evaluate the effectiveness of BO based black-box attacks using a spam email dataset. The attacker's objective is to create a spam email, $\mathbf{x}^*$, that is misclassified by the unknown classifier but is within $C$ of the original spam email $\mathbf{x}^A$.

**Spam Email Dataset** The dataset [19] contains 4601 records and each record holds 57 attributes. Among the 57 features, 2 of them are integers (we discard these two attributes as we currently only handle continuous features). Every email is labeled as either spam or normal. We randomly choose 3500 of the instances to train three different classifiers, and report the error rate on the remaining dataset. The original instance $\mathbf{x}^A$ is randomly selected from the spam emails testing set.

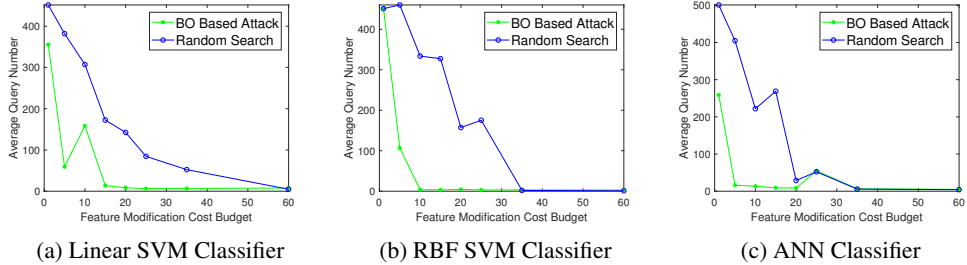

(a) Linear SVM Classifier  (b) RBF SVM Classifier  (c) ANN Classifier

Figure 1: Average Query Number w.r.t Different Cost Budgets for Different Classifiers

**Classifier Models** We train both linear and RBF kernel probabilistic SVM, which achieve classification accuracy of $91\%$ and $94\%$ respectively. Details of transforming normal SVM into probabilistic SVM can be found in Platt [20]. We also train an ANN model with classification accuracy of $94\%$.

**Results and Discussion** We compare our results with random search, which randomly generates values for each dimension and terminates the search process when the class label is changed. Specifically, we take the cost budget $C$ and generate random samples whose $L_1$-distance to $\mathbf{x}^A$ is in the range of $(C-\epsilon, C)$. We set $\epsilon = 0.05$. Our assumption here is, having larger distance to the original instance can maximize the chance of flipping into opponent class, assuming the boundary of the classifier does not have a highly irregular shape.

For different classifiers, we compare the number of queries needed to find the first successful adversarial example for both algorithms (BO attack and random search) as we vary $C$ from 1 to 60. Note that, when $C$ is extremely small, the chance of finding an adversarial example within the boundary is rare. Hence, we set some threshold values for both algorithms and once the iteration number exceeds the threshold, we take it as an indicator of non-existence of adversarial example. For our experiments, we set 50 as the upper limit for BO attack, and 500 for random search.

Figure 1 shows the average query number with respect to different feature modification cost budget $C$ for the three models. We see significant reductions in the number of queries using the BO attack for all of the classifiers. Note that, the average number of queries shown here is a conservative estimate for the BO method, since we take all iterations of BO exceeding 50 as failure and compute the iteration number as 500 for fair comparison with random search method. For cases where $C$ is small, our BO attack is substantially more efficient. For example, for the ANN Classifier with $C = 10$, it takes 16 queries to find the first successful adversarial example, compared to over 400 for the random search.

## 5 Conclusion

Our proposed black-box attack strategy considers the problem of generating adversarial example with minimum number of queries. We believe there are many scenarios where attackers will be limited in the number of interactions they have with a target model, so understanding the number of queries needed to find adversarial examples with high probability is an important problem. Our proposed Bayesian optimization approach shows promise in our preliminary experiments, and it is a general strategy that can be used against any classifier.

Our ongoing work will improve the BO attacks, compare with other black-box attacking methods, and test on data from different domains (e.g., image and text). We also note that, our approach can work for both targeted and untargeted attack. For targeted attacks, we simply set the objective function as maximizing $\Pr[f(\mathbf{x}) == y^*]$, where $y^*$ is the target class.

It is currently under investigation if the current approach can scale up to very high dimensions (e.g., image dataset) as performance of Bayesian optimization typically deteriorates in high dimensional case. For future directions, it is interesting to study the problem in a more restricted black-box case, where attackers only get classification labels. Another interesting direction is to provide a provable minimum modification to the original instance for adversarial sample generation in black-box setting.